\documentclass[sip,biber]{now-journal} 

\journaltitle{}

\usepackage[utf8]{inputenc}
\usepackage{amssymb}
\usepackage{float}

\usepackage{amsthm,amsfonts}
\usepackage{mathtools}
\usepackage{mathrsfs}
\usepackage{mdframed}
\usepackage{siunitx}
\usepackage{xurl}
\usepackage{algorithm}
\usepackage{algpseudocode}
\usepackage{booktabs,multirow}

\DeclareMathOperator*{\argmax}{arg\,max}

\DeclareMathOperator{\sign}{sign}
\newcommand*{\xcorr}{\star}

\newcommand*{\placeholder}{\mathinner{\cdot}}

\theoremstyle{plain}
\newtheorem{remark}{Remark}

\title{On the Invariance of Cross-Correlation Peak Positions Under Monotonic Signal Transformations, with Application to Fast Time Difference Estimation}

\author[1*]{Ueno,Natsuki}
\author[2]{Sato,Ryotaro}
\author[2]{Ono,Nobutaka}

\affil[1]{Faculty of Advanced Science and Technology, Kumamoto University}
\affil[2]{Department of Computer Science, Graduate School of Systems Design, Tokyo Metropolitan University}


\creditline*{Corresponding author: natsuki.ueno@ieee.org}

\begin{document}

\begin{abstract}
  We present a theorem concerning the invariance of cross-correlation peak positions.
  This theoretical result provides the foundation for a new method for time difference estimation that is potentially faster than the conventional fast Fourier transform (FFT) approach for real/complex sequences.
  Specifically, it shows that the peak position of the cross-correlation function between two shifted discrete-time signals remains unchanged under arbitrary monotonic transformations of the input signals.
  By exploiting this property, we design an efficient estimation algorithm based on the cross-correlation function between signals quantized into low-bit integers.
  The proposed method requires only integer arithmetic instead of real-valued operations, and further computational efficiency can be achieved through number-theoretic algorithms.
  Numerical experiments demonstrate that the proposed method achieves a shorter processing time than conventional FFT-based approaches within a specific range of signal lengths.
\end{abstract}

\section{Introduction}

The estimation of the time difference between two shifted time-series signals~\cite{Knapp:IEEE1976, Azaria:IEEE1984, Yamaoka:IEEE2023} is a fundamental task in many areas of audio signal processing. 
Typical applications include the synchronization of signals~\cite{Kammerl:ICASSP2014, Cabot:MOA1997, Chinaev:IEEE2021, Wang:IEEE2016}, the detection of a specific pattern in audio signals~\cite{Martinez:Adhoc2014}, and various array signal processing techniques such as source/sensor localization and tracking~\cite{Wax:IEEE1983, Doclo:EURASIP2003, Le:IEEE2017}.

One of the most classical yet effective approaches to time difference estimation is the peak detection of the \emph{cross-correlation function}~\cite{Carter:IEEE1987}. 
This approach has the advantage of its versatility, as it does not require special assumptions or prior knowledge on target signals. 
Additionally, the cross-correlation function can be efficiently calculated using fast Fourier transform (FFT) algorithms~\cite{Duhamel:Elsevier1990}, similarly to convolution. 
The Cooley--Tukey FFT algorithm~\cite{Cooley:MCOM1965, Burrus:IEEE1981, Duhamel:1984} is one of the most widely used FFT algorithms, and its core concept is still relevant today. 
For two signals of significantly different lengths, the overlap-add (or overlap-save) method~\cite{Stockham:AFIPS1966, Wefers:DAFx2011} is also effective when used with the FFT algorithms. 
Although there have been several improvements~\cite{Duhamel:Elsevier1990}, a further essential breakthrough in computational efficiency beyond these sophisticated strategies has yet to be realized. 

\begin{figure*}
  \centering 
  \includegraphics[width = 0.99\linewidth]{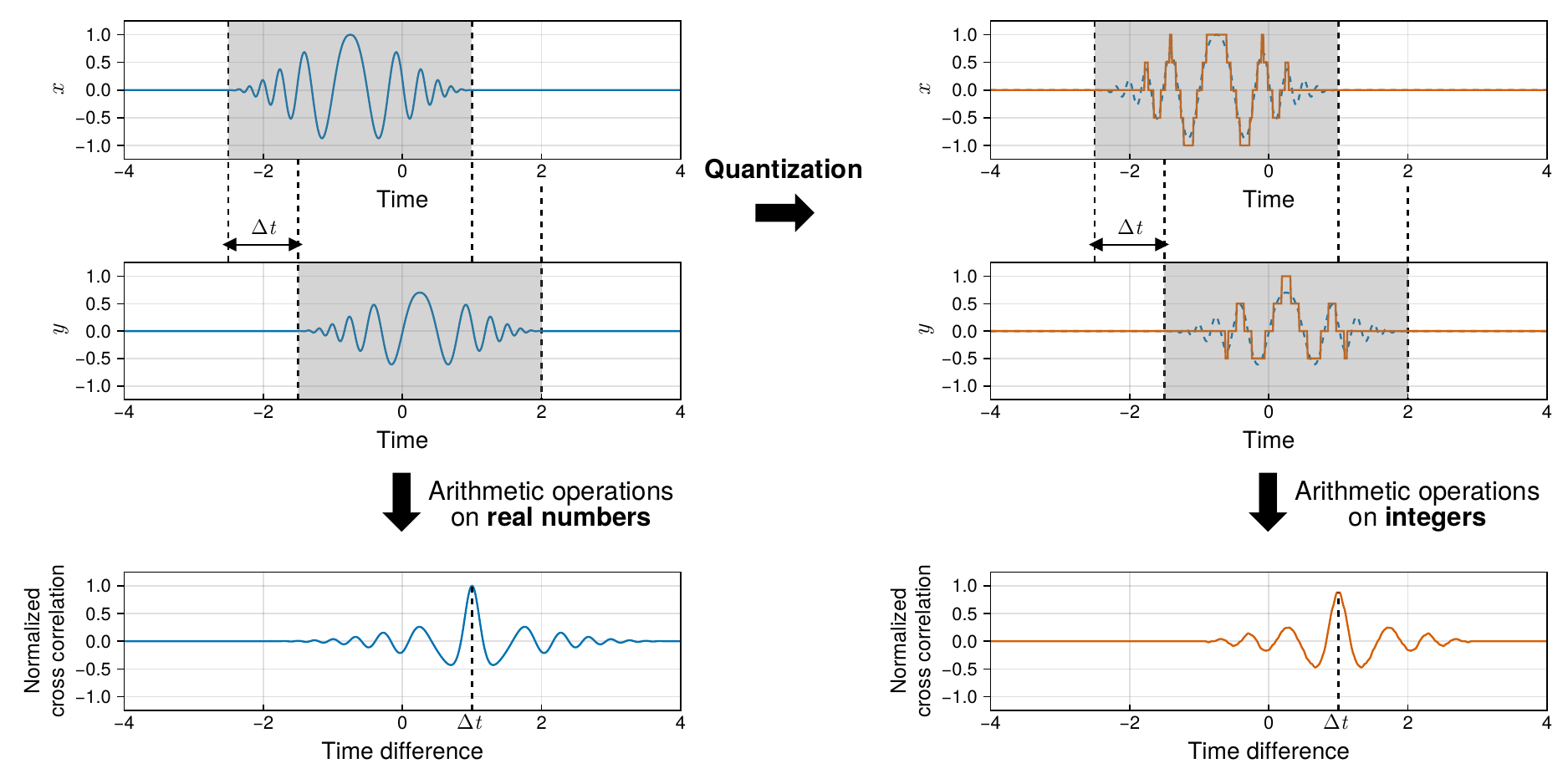}
  \caption{Overview of the proposed framework. Each graph on the third row represents the normalized cross-correlation function between $x$ and $y$, i.e., $(x\star y)/(\|x\|\cdot\|y\|)$.}
  \label{fig:overview}
\end{figure*}

Motivated by this background, this paper presents a new theorem useful for detecting the peak positions of cross-correlation functions and proposes a fast time difference estimation method leveraging this theorem.
Our theoretical result, derived from the \emph{rearrangement inequality}~\cite{Hardy:Cambridge1952}, shows that the peak position of the cross-correlation function between two shifted signals remains unchanged under any monotonic transformations applied to these two signals.
Consequently, we can estimate the time difference between two shifted signals by maximizing the cross-correlation function of these signals, which have been quantized into low-bit integers, as shown in Fig.~\ref{fig:overview}. 
This cross-correlation-like sequence can be computed only with arithmetic operations on integers instead of real/complex numbers, which significantly reduces computational complexity. 
Moreover, several number-theoretic algorithms can be employed for even more efficient computation~\cite{Furer:SIAM2009, Gathen:Cambridge2013, Harvey:AnnMath2021}. 
The validity of our approach was evaluated through experiments, the results of which demonstrated that the proposed method achieved a higher computational speed than the conventional method using the real/complex FFT algorithm within a specific range of signal lengths. 

\section{Notations and Preliminaries}

\subsection{Notations}

The sets of natural numbers excluding zero, integers, real numbers, and complex numbers are denoted by $\mathbb{N}$, $\mathbb{Z}$, $\mathbb{R}$, and $\mathbb{C}$, respectively. 
For a square-summable sequence $x: \mathbb{Z} \to \mathbb{R}$, $\|x\|$ denotes the $\ell_2$ norm of $x$.

\subsection{Formulation of Time Difference Estimation Problem}

Let $x, y: \mathbb{Z} \to \mathbb{R}$ be real-valued discrete-time signals with finite lengths, meaning they have only a finite number of non-zero values. 
The cross-correlation function between $x$ and $y$, denoted by $x \xcorr y: \mathbb{Z} \to \mathbb{R}$, is defined as 
\begin{equation}
  (x \xcorr y)[n] 
  \coloneqq 
  \sum_{m \in \mathbb{Z}} 
  x[m] \cdot y[m + n]
  \quad (n \in \mathbb{Z}). 
  \label{def:CCF} 
\end{equation}
Since $x$ and $y$ have finite lengths, the sum on the right-hand side of \eqref{def:CCF} always converges, and $x \xcorr y$ also has a finite length. 

As is well known, the time difference between two shifted signals can be estimated by maximizing their cross-correlation function. 
In mathematical terms, if $x$ and $y$ satisfy 
\begin{equation}
  x[n] = a \cdot y[n + \nu] \quad \forall n \in \mathbb{Z} 
  \label{eq:condition_theorem}
\end{equation}
for some $\nu \in \mathbb{Z}$ and $a \in (0,\infty)$, their cross-correlation function is maximized at $\nu$, i.e., 
\begin{equation}
  (x \xcorr y)[n] \leq (x \xcorr y)[\nu] \quad \forall n \in \mathbb{Z}. 
  \label{eq:inequality_CCF}
\end{equation}
This property follows immediately from the Cauchy--Schwarz inequality and the unitarity of the time shift operation. 

\subsection{Conventional Algorithms Using Real/Complex FFT}
\label{sec:conventional_algorithm}

When $x$ and $y$ have a length of $N \in \mathbb{N}$, meaning their supports are within a range of $N$ consecutive integers, the naive computation of $x \xcorr y$ requires $\Theta(N^2)$ multiplications on $\mathbb{R}$, where $\Theta(\placeholder)$ denotes the big-Theta notation. 
However, this computational complexity can be reduced by using FFT algorithms with zero padding. 
For example, the Cooley--Tukey FFT algorithm~\cite{Duhamel:Elsevier1990, Cooley:MCOM1965} allows the computation of $x \xcorr y$ with $\Theta(N \log N)$ multiplications on $\mathbb{C}$ (or equivalently on $\mathbb{R}$). 
Additionally, we also mention the prime-factor FFT algorithm~\cite{Burrus:IEEE1981}, which utilizes the Chinese remainder theorem, as another type of FFT algorithm, although it can be used in combination with the Cooley--Tukey FFT algorithm. 

\section{New Theorem and Its Application to Fast Time Difference Estimation}

\subsection{Theoretical Result}
\label{sec:mathematic}
Our main result, which forms the basis for fast time difference estimation, is summarized as follows.
\begin{theorem} 
  Let two finite-length signals $x,y: \mathbb{Z} \to \mathbb{R}$ satisfy \eqref{eq:condition_theorem} for some $\nu \in \mathbb{Z}$ and $a \in (0, \infty)$. 
  Then, for any monotonically non-decreasing functions $\varphi, \psi: \mathbb{R} \to \mathbb{R}$ satisfying $\varphi(0) = \psi(0) = 0$, the cross-correlation function between $\varphi \circ x$ and $\psi \circ y$ is maximized at $\nu$, i.e., 
  \begin{equation}
    ((\varphi \circ x) \xcorr (\psi \circ y))[n] 
    \leq 
    ((\varphi \circ x) \xcorr (\psi \circ y))[\nu] 
    \quad \forall n \in \mathbb{Z}.
  \end{equation}
  \label{thm:main}
\end{theorem}
\begin{remark}
  This theorem does not guarantee an order-preserving relationship between $(\varphi \circ x) \xcorr (\psi \circ y)$ and $x \xcorr y$. 
  Their elements are not necessarily in the same order except at the peak position, as illustrated in Fig.~\ref{fig:overview}. Nevertheless, both functions are maximized at the same position $\nu$. 
  As can also be seen from Fig.~\ref{fig:overview}, $\varphi \circ x$ and $\psi \circ y$ are no longer in a simple time-shift relationship; nevertheless, the peak positions of their cross-correlation functions remain invariant. 
  It is also worth noting that $(\varphi \circ x) \xcorr (\psi \circ y)$ can exhibit multiple peaks, although always including $\nu$, even when $x \xcorr y$ has only a single peak. 
\end{remark}
\begin{proof} 
  For each fixed $n \in \mathbb{Z}$, let $\Lambda_n \subset \mathbb{Z}$ be the index set of all $m \in \mathbb{Z}$ such that $x[m] \neq 0$ or $y[m+n] \neq 0$. 
  Then, by definition, we have 
  \begin{align}
    ((\varphi \circ x) \xcorr (\psi \circ y))[n] 
    = 
    \sum_{m \in \Lambda_n} (\varphi \circ x)[m] \cdot (\psi \circ y)[m + n]. 
    \label{eq:proof_1}
  \end{align}
  In addition, let $M_n \in \mathbb{N}$ be the number of elements in $\Lambda_n$, and $u_n, v_n: \{1, \ldots, M_n\} \to \mathbb{R}$ be respectively the descending sorts of $(x[m])_{m \in \Lambda_n}, (y[m + n])_{m \in \Lambda_n}$. 
  Then, since $\varphi \circ u_n$ and $\psi \circ v_n$ are also the descending sorts of $((\varphi \circ x)[m])_{m \in \Lambda_n}$ and $((\psi \circ y)[m + n])_{m \in \Lambda_n}$, respectively, we have 
  \begin{align}
    & \sum_{m \in \Lambda_n} 
    (\varphi \circ x)[m] \cdot (\psi \circ y)[m + n]
    \nonumber \\ 
    & \quad \leq 
    \sum_{m = 1}^{M_n} (\varphi \circ u_n)[m] \cdot (\psi \circ v_n)[m] 
    \label{eq:proof_2}
  \end{align}
  from the rearrangement inequality (Theorem 368 in \cite{Hardy:Cambridge1952}).
  On the other hand, from \eqref{eq:condition_theorem} and the monotonically non-decreasing properties of $\varphi$ and $\psi$, the two sequences $((\varphi \circ x)[m])_{m \in \Lambda_n}$ and $((\psi \circ y)[m + \nu])_{m \in \Lambda_n}$ are aligned in the same order. 
  Moreover, since $\varphi(0) = \psi(0) = 0$, these sequences include all non-zero elements of $\varphi \circ x$ and $\psi \circ y$.
  Therefore, we have 
  \begin{align}
    ((\varphi \circ x) \xcorr (\psi \circ y))[\nu] 
    & = 
    \sum_{m \in \Lambda_n} 
    (\varphi \circ x)[m] \cdot (\psi \circ y)[m + \nu]
    \nonumber\\
    & = 
    \sum_{m = 1}^{M_n} (\varphi \circ u_n)[m] \cdot (\psi \circ v_n)[m]. 
    \label{eq:proof_3}
  \end{align}
  Finally, from \eqref{eq:proof_1}--\eqref{eq:proof_3}, the inequality 
  \begin{equation}
    ((\varphi \circ x) \xcorr (\psi \circ y))[n] 
    \leq 
    ((\varphi \circ x) \xcorr (\psi \circ y))[\nu] 
  \end{equation}
  holds for any $n \in \mathbb{Z}$. 
\end{proof}

This theorem implies that we can estimate the time difference between two shifted signals, $x$ and $y$, by detecting the peak of the sequence $(\varphi \circ x) \xcorr (\psi \circ y)$ instead of $x \xcorr y$. 
Here, by applying integer-valued quantization as $\varphi$ and $\psi$, the computation of $(\varphi \circ x) \xcorr (\psi \circ y)$ requires only arithmetic operations on integers, rather than on real numbers as in the computation of $x \xcorr y$.

\subsection{Proposed Fast Time Difference Estimation Algorithm}
\label{sec:proposed_algorithm}

\begin{algorithm}[t]
  \caption{Proposed time difference estimation algorithm}
  \label{alg:proposed}
  \begin{algorithmic}[1]
    \Require{$x, y: \mathbb{Z} \to \mathbb{R}$ of finite lengths}
    \State Set monotonically non-decreasing functions $\varphi, \psi : \mathbb{R} \to \mathbb{Z}$ satisfying $\varphi(0) = \psi(0) = 0$
    \State $u, v \gets \varphi \circ x, \psi \circ y$
    \State $w \gets u \xcorr v$
    \State $\Lambda \gets \argmax_{\nu \in \mathbb{Z}} (w[\nu])$
    \If{$\Lambda$ has multiple indices} 
    \State $\Lambda \gets \argmax_{\nu \in \Lambda} (x \xcorr y)[\nu]$
    \EndIf 
    \State \Return $\Lambda$
  \end{algorithmic}
\end{algorithm}

Our proposed time difference estimation algorithm utilizing the aforementioned idea is summarized in Algorithm~\ref{alg:proposed}. 
Any quantization scheme, including nonuniform quantization, can be applied as $\varphi$ and $\psi$ as long as their ranges are included within $\mathbb{Z}$. 
Under the conditions imposed in Theorem~\ref{thm:main}, it is guaranteed that Algorithm~\ref{alg:proposed} precisely obtains the index that maximizes $x \star y$. 
On the other hand, when the signals contain noise, an inherent trade-off exists between the quantization depth and estimation accuracy.
However, it will be demonstrated by the experiments in Section~\ref{sec:experiments_accuracy} that even with extreme quantization using the sign function, the time difference can still be accurately estimated in practical situations. 

Unless excessively many peaks are detected in the fifth line, the dominant computational complexity of Algorithm~\ref{alg:proposed} arises in the third line, i.e., in the calculation of the cross-correlation function between two finite-length integer-valued sequences. 
Unfortunately, the fast computation strategies employed in real/complex FFT algorithms, as described in Section~\ref{sec:conventional_algorithm}, are not directly applicable to integer-valued sequences. 
Instead, several number-theoretic algorithms for multiplying univariate polynomials over a ring of integers modulo a fixed integer are available. 
This is because the cross-correlation function between two signals corresponds to the convolution between one signal and the time-reversed version of the other signal, and convolution is essentially equivalent to polynomial multiplication. 
Representative algorithms include the Kronecker substitution, the Karatsuba algorithm, the Sch\"{o}nhage--Strassen algorithm, and the multimodular method based on the Chinese remainder theorem~\cite{Gathen:Cambridge2013, Harvey:Elsevier2009, Harvey:AnnMath2021, Weimerskirch:IACR2006}. 
As an alternative approach, it is also possible to compute convolutions approximately using the integer FFT~\cite{Oraintara:IEEE2002, Rao:Springer2010}; however, even under noise-free conditions, exact detection of peak positions cannot be strictly guaranteed.

A unified analysis of the computational complexity of the proposed framework is challenging owing to the variations in the procedure used and computational complexity among the aforementioned approaches. 
Here, we outline the computation method based on Kronecker substitution as an example, along with its associated computational complexity. 
For simplicity, consider the product of two univariate polynomials, $U$ and $V$, defined as 
\begin{equation}
  U(p) = \sum_{n = 0}^{N-1} u[n] \cdot p^n, \quad
  V(p) = \sum_{n = 0}^{N-1} v[n] \cdot p^n,
\end{equation}
where the coefficients are given by integer-valued sequences $u,v: \{0, \ldots, N-1\} \to \{-K, \ldots, K\}$ of length $N \in \mathbb{N}$, with an absolute value of at most $K \in \mathbb{N}$. 
Using the convolution operator $\ast$, their product $U \cdot V$ is given by 
\begin{equation}
  (U \cdot V)(p)
  = \sum_{n = 0}^{2N - 2} (u \ast v)[n] \cdot p^n,
\end{equation}
meaning the essential equivalence between the product of polynomials and the convolution of sequences. 
Since $-NK^2 \leq (u \ast v)[n] \leq NK^2$ for all $n \in \{0, \ldots, 2N-2\}$, the coefficients of $U \cdot V$, i.e., the sequence $u \ast v$, can be determined uniquely from the value of
\begin{equation}
  U(2^L) \cdot V(2^L)
  = 
  \sum_{n = 0}^{2N - 2} (u \ast v)[n] \cdot 2^{Ln}
\end{equation}
by using $L \in \mathbb{N}$ satisfying $2^L > 2NK^2$. 
For a specific example, refer to \cite{Harvey:Elsevier2009}. 
Here, since $U(2^L)$ and $V(2^L)$ are $NL$-bit signed integers, their multiplication can be computed with \emph{bit operations} of complexity $\Theta(NL \log (NL) \log \log (NL))$ using the Sch\"{o}nhage--Strassen algorithm, a standard method for integer multiplication\footnote{Fast algorithms for integer multiplication remain an active research area, and it has been shown that the multiplication of two $N$-bit integers can be achieved with a computational complexity of $O(N \log N)$~\cite{Harvey:AnnMath2021}, where $O(\placeholder)$ denotes the big-O notation. However, in practical applications, the Sch\"{o}nhage--Strassen algorithm is commonly used.}. 
Note that the calculation of $U(2^L)$ and $V(2^L)$ from $u$ and $v$ and the reconstruction of $u \ast v$ from the value of $U(2^L) \cdot V(2^L)$ are not dominant compared with the integer multiplication mentioned above. 
By optimally choosing $L = \lfloor \log_2 NK^2 \rfloor + 2$, the computational complexity with respect to $N$ becomes $\Theta(N (\log N)^2 \log \log N)$. 
Thus, although the asymptotic computational complexity with respect to the signal length $N$ is smaller for the real/complex FFT, the proposed method is expected to improve computation speed when $N$ is not excessively large because the proposed method replaces arithmetic operations on $\mathbb{R}$ or $\mathbb{C}$ with bit operations. 

\section{Experiments}
\label{sec:experiments}

\subsection{Comparison of Computation Time}

First, to evaluate the effectiveness of the proposed approach, we compared the processing time under the following three conditions.
\begin{itemize}
  \item \textbf{Integer-KS}: calculation of cross-correlation function of integer-valued sequences with the Kronecker substitution.
  \item \textbf{Real-Naive}: naive $\Theta(N^2)$ computation of the cross-correlation function for real-valued (double-precision) sequences.
  \item \textbf{Real-FFT}: calculation of cross-correlation function of real-valued (double-precision) sequences with the FFT algorithm.
\end{itemize}
The lengths of the two signals were set equally to $N$, and the processing time was measured for different values of $N$. 
Here, the two signals were randomly generated since the focus of this experiment was solely on computation time.
The evaluation program was written in C++ and compiled using GCC 13.3.0 with the -O2 optimization option. 
The CPU used for evaluation was Intel\textsuperscript{\textregistered} Core\textsuperscript{\texttrademark} Ultra 9 Processor 285K with $\SI{64}{GB}$ of RAM. 
In \textbf{Integer-KS}, the GMP library was used for integer multiplications. 
In \textbf{Real-FFT}, the FFTW library~\cite{Frigo:IEEE2005} was used with the \texttt{FFTW\_MEASURE} option. 
Here, an optimized implementation that removes the symmetric redundancy of the discrete Fourier transform resulting from the real-valued input signals was used for fast computation. 
Under all conditions, the processing time was evaluated as the average of $\max\{16, 2^{16}/N\}$ trials, excluding the first trial to eliminate the effect of planning time in the FFT. 

\begin{figure}
  \centering
  \includegraphics[width = 0.8\linewidth]{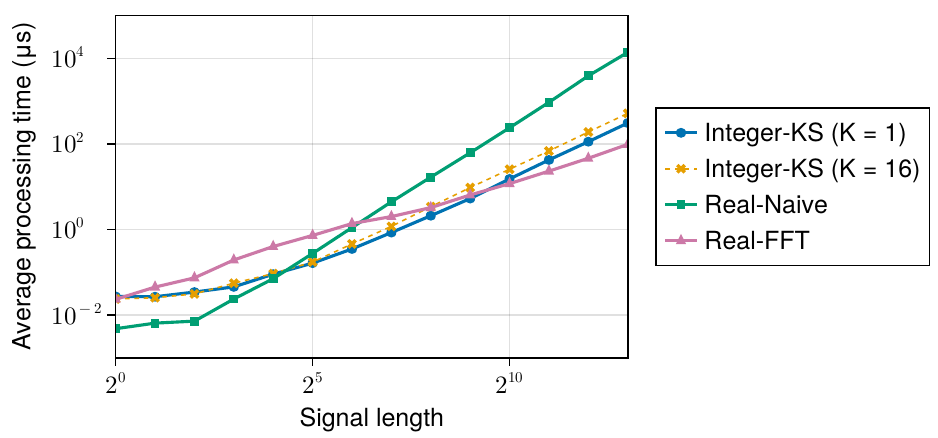}
  \caption{Average processing time plotted against signal length.}
  \label{fig:computation_time}
\end{figure}

Figure~\ref{fig:computation_time} shows the relationship between the signal length and the average processing time. 
It can be observed that \textbf{Integer-KS} achieved faster computation than \textbf{Real-FFT} for small $N$, which is consistent with the discussion in Section~\ref{sec:proposed_algorithm}. 
Furthermore, it was confirmed that the use of the Kronecker substitution in \textbf{Integer-KS} enabled even faster computation than that in \textbf{Real-Naive} for large $N$. 
As a concrete result, when $K=1$, \textbf{Integer-KS} achieved faster processing than both \textbf{Real-Naive} and \textbf{Real-FFT} in the range $2^5 \leq N \leq 2^9$, suggesting the effectiveness of the proposed framework. 

\subsection{Evaluation of Robustness Against Background Noise}
\label{sec:experiments_accuracy}

We also investigated the impact of the quantization in the proposed method on the estimation accuracy under realistic noisy conditions. 
As a simple task, we evaluated the accuracy of time difference estimation between $x$ and $y$, where $y$ was the target signal and $x$ was the noisy signal, i.e., the mixture of the time-shifted target signal and the background signal.
The target signal was a one-second \SI{16}{kHz}/\SI{16}{bit} speech signal randomly drawn from 10 word categories (\textit{down}, \textit{go}, \textit{left}, \textit{no}, \textit{off}, \textit{on}, \textit{right}, \textit{stop}, \textit{up}, \textit{yes}; 100 samples per category) in the Speech Commands Dataset~\cite{Warden:arXiv2018}. 
The background signal was a five-second environmental audio signal from the ESC-50 Dataset~\cite{Piczak:MM2015}, which consists of 50 classes grouped into five major categories: animals, natural soundscapes and water sounds, human non-speech sounds, interior/domestic sounds, and exterior/urban noises. 
For the experiments, 1000 samples of background signals were constructed by stratified sampling (20 samples from each class), paired randomly with target signals, and resampled to \SI{16}{kHz} from the original sampling rate of \SI{44.1}{kHz}. 
The true time difference was determined according to the uniform distribution on $\SI{[0, 4]}{s}$, and the subsample shift was calculated using sinc interpolation.
The mixture weight was set so that the ratio of the squared $\ell_2$ norm per unit time of the target signal to that of the background signal corresponded to the given signal-to-noise ratio (SNR). 

The time difference was estimated in integer sample units using the proposed method given in Algorithm~\ref{alg:proposed} (denoted by \textbf{Proposed}) and the conventional method that maximizes $x \star y$ (denoted by \textbf{CCF wo/ quantization}). 
We also evaluated a method restricted to the first four lines of Algorithm 1 (denoted by \textbf{Proposed (until 4th line)}).
The sign function, with $\sign(0)$ defined as 0, was used for the quantization functions $\varphi$ and $\psi$, which corresponds to the case of $K = 1$ in Section~\ref{sec:proposed_algorithm}. 
A thousand pairs of $x$ and $y$ were evaluated, and in each trial, the estimation was regarded as correct when the true and estimated time differences $\nu$ and $\hat{\nu}$ in sample units satisfied $|\hat{\nu} - \nu| < 1$. 
Here, when multiple time differences were estimated, the estimation was regarded as correct only if all estimated time differences satisfied the above condition. 

\begin{figure}
  \centering
  \includegraphics[width = 0.65\linewidth]{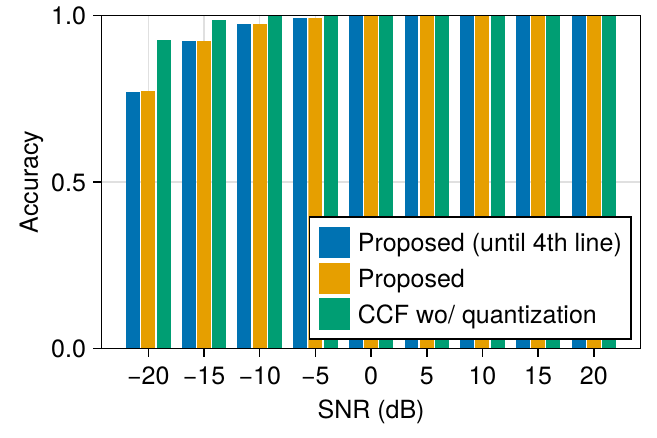}
  \caption{Estimation accuracy of the time difference plotted against SNR. \textbf{Proposed} and \textbf{Proposed (until 4th line)} were evaluated with $K=1$.}
  \label{fig:accuracy}
\end{figure}

\begin{figure}
  \centering
  \includegraphics[width = 0.9\linewidth]{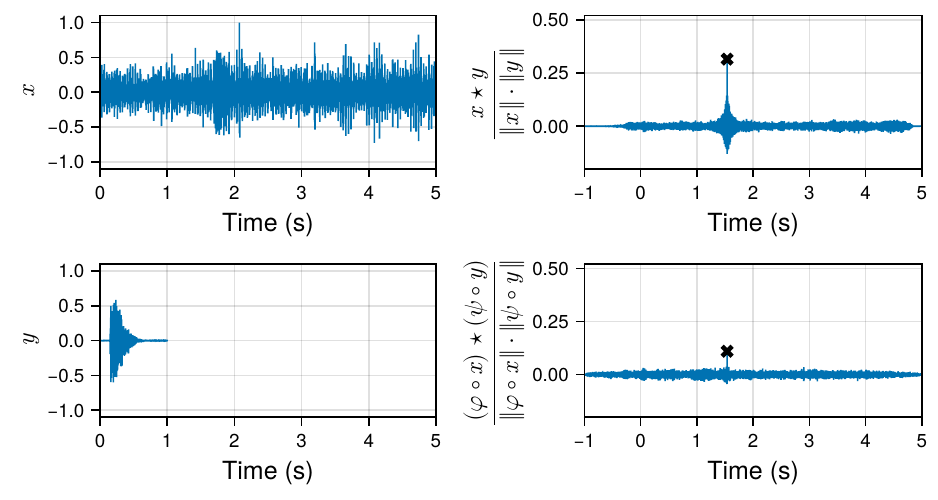}
  \caption{Noisy signal $x$ (top left); target signal $y$ (bottom left); the normalized cross-correlation function between $x$ and $y$ (top right); the normalized cross-correlation function between $\varphi \circ x$ and $\psi \circ y$ (bottom right). The cross symbols represent the peak positions of the cross-correlation functions.}
  \label{fig:example}
\end{figure}

Figure~\ref{fig:accuracy} shows the relationship between the SNR and estimation accuracy (number of correct trials per total number of trials).
Despite extreme quantization, the proposed method achieved nearly perfect estimation when the SNR was higher than $\SI{0}{dB}$. 
Furthermore, the comparison between \textbf{Proposed} and \textbf{Proposed (until 4th line)} implies that, in the proposed method, there is rarely a need to refer to the values of the original cross-correlation function.
An example of the estimation result for one trial at an SNR of $\SI{-10}{dB}$, where the target signal was an utterance of \textit{down} and the background signal was the sound of sea waves, is shown in Fig.~\ref{fig:example}. 
In this trial, the true time difference was $17005.90$ in sample units, and all three methods estimated the time difference as $17006$ in sample units. 
It can be observed that the maximum value of the cross-correlation function for \textbf{Proposed} (bottom right) is located at the same position as in the case of \textbf{CCF wo/ quantization} (top right), although the sharpness of the peak decreases owing to quantization.

\section{Conclusion}

We present a theorem useful for peak detection of the cross-correlation function between two shifted signals and apply this theorem to develop a fast algorithm for time difference estimation.
Experiments confirmed that the proposed method operates faster than the conventional FFT-based approach, with minimal degradation in estimation performance under practical conditions. 
Future work includes incorporating the overlap-add method into the proposed framework for long-signal and streaming scenarios, as well as investigating more efficient quantization schemes that better balance computational cost and estimation robustness.

\section*{Acknowledgements}

\begin{sloppypar}
This work was supported by JSPS KAKENHI Grant Number JP23K16904.
\end{sloppypar}

\section*{Code Availability}
The demo code of the proposed method and the reproducibility scripts for the experiments are publicly available at \url{https://github.com/natsuenono/fast-time-difference-estimation-demo} and \url{https://github.com/natsuenono/fast-time-difference-estimation-experiments}, respectively.

\printbibliography

\end{document}